\documentstyle[amsmath,amssymb,graphicx,multicol]{article}

\def\be{\begin{eqnarray}}
\def\ee{\end{eqnarray}}

\hoffset= - 1.0in         

\title{{\bf Interplay between MacDonald and Hall-Littlewood expansions
of extended torus superpolynomials  } \vspace{.2cm}}
\author{{\bf A.Mironov}\footnote{ {\small {\it
Lebedev Physics Institute} and {\it ITEP, Moscow, Russia}};
mironov@itep.ru; mironov@lpi.ru}, {\bf A.Morozov}\thanks{{\small
{\it ITEP, Moscow, Russia}}; morozov@itep.ru},
{\bf
Sh.Shakirov}\thanks{{\small {\it Department of Mathematics, University of
California, Berkeley, USA}, {\it Center for Theoretical Physics, University
of California, Berkeley, USA} and {\it
ITEP Moscow, Russia}}; shakirov@math.berkeley.edu}\ \  and
{\bf A.Sleptsov}\thanks{{\small
{\it ITEP, Moscow, Russia}};
sleptsov@itep.ru}\date{ }}

\begin{document}
 \maketitle

\vspace{-5.5cm}

\begin{center}
\hfill FIAN/TD-22/11\\
\hfill ITEP/TH-01/12\\
\end{center}

\vspace{5cm}

\centerline{ABSTRACT}

\bigskip

{\footnotesize In \cite{DMMSS} extended superpolynomials were
introduced for the torus links $T[m,mk+r]$, which are functions on
the entire space of time variables and, at expense of reducing the
topological invariance, possess additional algebraic properties,
resembling those of the matrix model partition functions and the
KP/Toda tau-functions. Not surprisingly, being a suitable extension
it actually allows one to calculate the superpolynomials. These
functions are defined as expansions into MacDonald polynomials, and
their dependence on $k$ is entirely captured by the action of the
cut-and-join operator, like in the HOMFLY case. We suggest a simple
description of the coefficients in these character expansions, by
expanding the initial (at $k=0$) conditions for the $k$-evolution
into the new auxiliary basis, this time provided by the
Hall-Littlewood polynomials, which, hence, play a role in the
description of the dual $m$-evolution. For illustration we list
manifest expressions for a few first series, $mk\pm 1, mk\pm 2,
mk\pm 3, mk\pm 4$. Actually all formulas were explicitly tested up to $m=17$
strands in the braid.}

\vspace{1.5cm}

\section{Introduction}

In \cite{DMMSS} a general expression was suggested for
the superpolynomials \cite{Kh,superpols} of the torus knots and links,
which is actually a W-representation of \cite{MMN} (ultimately related with matrix
model representations \cite{Wrep} and
Hurwitz theory \cite{MMN}),
generalizing the known expression of this kind \cite{chi,knMMM12} for
the torus HOMFLY polynomials,
of which it is actually a $\beta$-deformation \cite{betadefo}
with $t=q^\beta$.
It directly reproduces all available answers for particular
torus knots obtained by several alternative methods in
\cite{EG,Obl,ASh,Che,ShSh}.
It is naturally generalized to the {\it colored} superpolynomials,
but, in this letter, we restrict consideration to the case of
the first fundamental representation $R=\Box=[1]$.

The main idea is to extend superpolynomials to  functions of
a $\tau$-function type, depending on infinitely many time variables $p_k$,
which are no longer knot invariants (they depend on the braid
representation of the knot), but instead is a nice algebraic
quantity, possessing a variety of hidden symmetries.
It has natural form of a character decomposition,
which, in the $\beta$-deformed case, is the decomposition into
the MacDonald polynomials
$M_Q\{p\}$,
\be
\boxed{\ \
{\cal P}_{m,mk+r}\{p\}
{\phantom{{{{{5^5}^5}^5}^5}^5}} = {\phantom{{{{{5^5}^5}^5}^5}^5}}
\sum_{Q\vdash m} c_Q^{(m,r)} q^{-k\nu(Q^T)}t^{k\nu(Q)} M_Q\{p\}
= e^{k\hat W} {\cal P}_{m,r}\{p\}\ \
}
\label{tspexpan}
\ee
where $[m,n]=[m,mk+r]$, $\ 0\leq r< m\ $ dentes the torus link,
$Q = \{Q_1\geq Q_2\geq\ldots\geq Q_{l(Q)}>0\}$ are the Young diagrams
of the size $\sum_{i=1}^{l(Q)} Q_i = m$ with $l(Q)$ rows,
$Q^T$ is the transposed diagram of $Q$,
and $\nu(Q) = \sum_{i=1}^{l(Q)}(i-1)Q_i$.
Both $M_Q$ and $c_Q$ also depend on the two deformation parameters $q$
and $t=q^\beta$.

This ${\cal P}_{m,n}\{p\}$ reduces to an ordinary superpolynomial
$P_{m,n}(A) = {\cal P}_{m,n}\{p=p^*\}  = {\cal P}_{m,n}^*$
at a special $1d$ locus
in the space of time variables, parameterized by $A = t^N$,
\be
p_k = p^*_k = \frac{1-A^k}{1-t^k} = [N]_{t^k}
\ee
It becomes a polynomial with all coefficients positive in the
case of knot (for $m$ and $n$ coprime) and after being expressed
in the special variables\footnote{
In this letter we use the notation with "asymmetric" quantum numbers,
$[x]_t = \frac{1-t^x}{1-t}$, which looks most adequate for torus knots.
This is different from \cite{DMMSS}, where the symmetric choice was
made, with $[x]_t^{\cite{DMMSS}} = \frac{t^x-t^{-x}}{t-t^{-1}}$,
differing in particular by the change $t\rightarrow t^2$, $q\rightarrow q^2$.
The symmetric choice is good, it eliminates artificial square roots
from formulas for generic knots. However, because of an additional $Z_2$-symmetry
in the torus case, the asymmetric notation provides simplifications,
and we use it in this short letter to make formulas as simple as possible.
}
\be
{\bf t} = -\sqrt{q/t}, \ \ \ {\bf q} = \sqrt{t}, \ \ \ {\bf a}^2 = A\sqrt{t/q}
\ee

A non-trivial part of the story is a description of the expansion
coefficients $c_Q$.
For the HOMFLY polynomials these were essentially
$q$-independent integers,
read off from the Adams decomposition at the "initial" point $n=0$ \cite{chi}.
After $\beta$-deformation they become non-trivial rational
functions of $q$ and $t$, and are {\it not} given by just a
naive deformation of the Adams rule \cite{DMMSS}.
As explicitly stated in (\ref{tspexpan}) they
do not depend on the "evolution" parameter $k$, only on the "series"
labeled by the residue $r = n\ {\rm mod}\ m$.
These coefficients are straightforwardly calculated, by using
\begin{itemize}
\item the duality ${\cal P}^*_{m,n}(A) \sim A^{m-n}{\cal P}_{n,m}^*(A)$
at the "initial" point $k=0$, i.e. $n=r$: it provides a recurrent relation
in $m$, allowing one to go down from $m$ to $m'=r<m$, and
\item the lifting rules, allowing to continue the superpolynomial
at the "initial" point $k=0$ from the locus $\{p^*\}$ to the entire $\{p\}$-space;
\item another initial condition at $k=-1$ can be used to additionally test the
results; to do so one should use the symmetries
$T[m,-n] \leftrightarrow T[m,n]$
and $T[m,r] \leftrightarrow T[m,m-r]$, see formula (\ref{dual}).
\end{itemize}
All this is explained and illustrated in great detail in \cite{DMMSS},
and an important problem is to find a convenient description of
the rather sophisticated combinatorial functions $c_Q^{(m,r)}$.
In this letter, we suggest a possibility which looks very promising.
The key observation is that the extended superpolynomial (\ref{tspexpan})
at $k=0$ has a nice decomposition in terms of the Hall-Littlewood polynomials
$L_Q\{p\} = \left.M_Q\{p\}\right|_{q=0}$:
\be\label{dec}
{\cal P}_{m,r}\{p\} = \sum_{Q\vdash m} c_Q^{(m,r)}M_Q\{p\}
= \sum_{\stackrel{Q\vdash m}{l(Q)\leq r}} h_Q^{(m,r)} L_Q\{p\}
\ee
namely the coefficients $h_Q$ are non-vanishing only for $l(Q)\leq r$
(no more than $r$ rows)
and possess additional algebraic properties.
We list now the generating function for these coefficients
for a few lowest values of $r$ to illustrate this statement.

\section{ Hall-Littlewood coefficients $h_Q$ for torus knots}

The coefficients $h_Q$ are in many ways simpler than the coefficients $c_Q$.
In particular, in variance with $c_Q$, the coefficients $h_Q$ are \emph{polynomials}
in $q,t$ with integer coefficients:
$$
h_Q \in {\mathbb Z}[q,t]
$$
It is convenient to separate a simple overall factor
$$
h_Q = q^{\nu(Q)} (1-t)^{l(Q)-1} \hat h_Q
$$
which makes $\hat h_Q$ normalized to unity: $\hat h_Q = 1 + O(q,t)$.

Let us give a list of explicit examples of these coefficients $\hat h_Q$, which
illustrates their properties. In the tables below, included are only the diagrams
for which the coefficients are non-vanishing. In the case of $r = 1$:

\paragraph{\underline{$(3,3k + 1)$:}}

\[
\begin{array}{c|llllll}
{\rm diagram} \ Q & {\rm coefficient} \ \hat h_Q & \rule{0pt}{3mm}  \\
\hline \
\big[3\big] & 1 & \rule{0pt}{3mm}  \\
\end{array}
\]

\paragraph{\underline{$(4,4k + 1)$:}}

\[
\begin{array}{c|llllll}
{\rm diagram} \ Q & {\rm coefficient} \ \hat h_Q & \rule{0pt}{3mm}  \\
\hline \
\big[4\big] & 1 & \rule{0pt}{3mm}  \\
\end{array}
\]

\paragraph{\underline{$(5,5k + 1)$:}}

\[
\begin{array}{c|llllll}
{\rm diagram} \ Q & {\rm coefficient} \ \hat h_Q & \rule{0pt}{3mm}  \\
\hline \
\big[5\big] & 1 & \rule{0pt}{3mm}  \\
\end{array}
\]

\noindent
and so on, i.e.

\be
\boxed{
\hat h_Q^{(m,1)} = \left\{ \begin{array}{ll} 1 & l(Q) = 1 \\
0 & {\rm otherwise} \end{array} \right.
}
\ee
The simplicity of this formula is stunning: it captures all the information
about the superpolynomials of $(m,mk+1)$ knots in a single unity.
As we will see below, from the point of view of the
Hall-Littlewood re-expansion, $(m,mk+1)$-series is by no means distinguished:
a similar phenomenon happens for the higher series, e.g. for
$(m,mk+2), (m,mk+3)$ etc.

\paragraph{\underline{$(3,3k + 2)$:}}

\[
\begin{array}{c|llllll}
{\rm diagram} \ Q & {\rm coefficient} \ \hat h_Q & \rule{0pt}{3mm}  \\
\hline \
\big[3\big] & 1 & \rule{0pt}{3mm}  \\
\big[2,1\big] & 1 & \rule{0pt}{3mm} \\
\end{array}
\]

\paragraph{\underline{$(5,5k + 2)$:}}

\[
\begin{array}{c|llllll}
{\rm diagram} \ Q & {\rm coefficient} \ \hat h_Q & \rule{0pt}{3mm}  \\
\hline \
\big[5\big] & 1 & \rule{0pt}{3mm}  \\
\big[4,1\big] & 1 & \rule{0pt}{3mm}  \\
\big[3,2\big] & 1 & \rule{0pt}{3mm}  \\
\end{array}
\]

\paragraph{\underline{$(7,7k + 2)$:}}

\[
\begin{array}{c|llllll}
{\rm diagram} \ Q & {\rm coefficient} \ \hat h_Q & \rule{0pt}{3mm}  \\
\hline \
\big[7\big] & 1 & \rule{0pt}{3mm}  \\
\big[6,1\big] & 1 & \rule{0pt}{3mm}  \\
\big[5,2\big] & 1 & \rule{0pt}{3mm}  \\
\big[4,3\big] & 1 & \rule{0pt}{3mm}  \\
\end{array}
\]

\noindent
and so on, i.e.

\be\label{h2}
\boxed{
\hat h_Q^{(m,2)} = \left\{ \begin{array}{ll}
1 & l(Q) = 1 \\
1 & l(Q) = 2 \\
0 & {\rm otherwise} \end{array} \right.
}
\ee
Remarkably, in these terms the case $mk+2$
looks just as simple as $mk+1$.
All complications arise when one
performs the linear transformation from  $h_Q^{(m.r)}$
to  $c_Q^{(m,r)}$.
In \cite{DMMSS} it was suggested to split $c_Q^{(m,r)}$
into the Adams coefficients and additional $\gamma$-factors,
which were trivial for the HOMFLY polynomials, i.e. for $q=t$:
$c_Q^{(m,r)} = C_Q^{(m,r)}\gamma_Q^{(m,r)}$.
Then for $r=1$ all the $\gamma$'s are simple polynomials,
$\gamma_Q^{(m,1)} \sim \sum_{(i,j)\in Q} t^iq^{-j}$
(we use here the elegant reformulation of this result from
\cite{DMMSS} suggested in \cite{ShSh}).
For $r=2$ these $\gamma$-factors are rather complicated
rational functions, for example,
\be
\gamma_{[4,3]}^{(7,2)} = q^4\
\dfrac{q^7 t^3+t^2 q^6-t^3 q^5+t q^5-q^4 t^3+q^4+2 q^4 t-q^4 t^2-2 q^3 t^2+
q^3+q^3 t-q^2 t^2-q^2 t+q^2-q t-1}{q^2-t}
\ee
while the corresponding Adams coefficient is
\be
C_{[4,3]}^{(7,2)} = \dfrac{(1-t)(q-t)(q^2-t)}{(1-q^2 t)(1-q^3 t)(1-q^4 t)}
\ee
But this seeming complexity is actually nothing, but
the result of the linear transformation from the Hall-Littlewood
to MacDonald polynomials, and is a direct consequence
of formulas (\ref{h2}).

Similarly, for $r = 3$,

\begin{multicols}{2}
\paragraph{\underline{$(4,4k + 3)$:}}

\[
\begin{array}{c|llllll}
{\rm diagram} \ Q & {\rm coefficient} \ \hat h_Q & \rule{0pt}{3mm}  \\
\hline \
\big[4\big] & 1 & \rule{0pt}{3mm}  \\
\big[3,1\big] & 1+q & \rule{0pt}{3mm} \\
\big[2,2\big] & 1+t & \rule{0pt}{3mm} \\
\big[2,1,1\big] & 1+t & \rule{0pt}{3mm} \\
\end{array}
\]

\paragraph{\underline{$(5,5k + 3)$:}}

\[
\begin{array}{c|llllll}
{\rm diagram} \ Q & {\rm coefficient} \ \hat h_Q & \rule{0pt}{3mm}  \\
\hline \
\big[5\big] & 1 & \rule{0pt}{3mm}  \\
\big[4,1\big] & 1+q & \rule{0pt}{3mm} \\
\big[3,2\big] & 1+q & \rule{0pt}{3mm} \\
\big[3,1,1\big] & 1+t & \rule{0pt}{3mm} \\
\big[2,2,1\big] & 1+t & \rule{0pt}{3mm} \\
\end{array}
\]

\paragraph{\underline{$(7,7k + 3)$:}}

\[
\begin{array}{c|llllll}
{\rm diagram} \ Q & {\rm coefficient} \ \hat h_Q & \rule{0pt}{3mm}  \\
\hline \
\big[7\big] & 1 & \rule{0pt}{3mm}  \\
\big[6,1\big] & 1+q & \rule{0pt}{3mm} \\
\big[5,2\big] & 1+q+q^2-qt & \rule{0pt}{3mm} \\
\big[4,3\big] & 1+q & \rule{0pt}{3mm} \\
\big[5,1,1\big] & 1+t & \rule{0pt}{3mm} \\
\big[4,2,1\big] & 1+q & \rule{0pt}{3mm} \\
\big[3,3,1\big] & 1+t & \rule{0pt}{3mm} \\
\big[3,2,2\big] & 1+t & \rule{0pt}{3mm} \\
\end{array}
\]
\end{multicols}

\noindent
and so on, i.e.
\be
\boxed{
\hat h_Q^{(m,3)} = \left\{ \begin{array}{ll}
1 & l(Q) = 1 \\
1 + t + (q-t)[\alpha]_q & l(Q) = 2,3 \\
0 & {\rm otherwise} \end{array} \right.
}
\ee
with $\alpha = \min(Q_1-Q_2,Q_2-Q_3)$. These examples clearly show the important
role of the Hall-Littlewood basis.

\section{ Generating functions }

As it often happens, the most adequate description of combinatorial information is
given by generating functions. In our case, the combinatorial objects under
consideration are the coefficients $\hat h_Q$. It is, therefore, convenient to pass
from explicit formulas for $\hat h_Q$, which depend on integer
variables $Q_1, Q_2, \ldots$, to generating functions which depend on
continuous variables $x_1, x_2, \ldots$ This is achieved by summing over all
diagrams $Q$ with appropriate weights $\omega_Q$:
\be
\check\rho_m(x_1, \ldots, x_r) = \sum_{|Q|=m} \omega_Q \hat h_Q \ x_1^{Q_1} \ldots x_r^{Q_r}
\ee
There is of course some freedom in the choice of the weights $\omega_Q$: it can be
used to simplify the generating functions as much as possible. In our case,
the weight that gives the simplest answer comes from the Hall-Littlewood theory,
it is essentially the inverse quadratic norm of the Hall-Littlewood polynomials:
\be
\omega_Q = \prod\limits_{j} \dfrac{1}{[m_j(Q)]_t!} = (1-t)^{l(Y)}||L_Q||^2
\ee
where $m_j(Y)$ = the number of rows with length $j$ in the diagram $Y$, and the quantum
factorial is defined as $[x]_t! = [1]_t \ldots [x]_t$. This is the norm which also appears in the Cauchy
formula
\be
\exp\left(\sum_k{z^k\over k}(1-t^k)p_k\bar p_k\right)=\sum_R {z^{|R|}\over ||L_Q||^2}
L_R\{p\}L_R\{\bar p\}
\ee

It is further convenient
to sum over all indices $m$ coprime with $r$:
\begin{align}
\rho(x_1, \ldots, x_r | z ) = \sum\limits_{m\perp r} z^m \check\rho_m(x_1, \ldots, x_r)
\end{align}
With these conventions, the 1-point function becomes
\begin{align}
\rho(x | z ) = \dfrac{x z}{1 - x z}
\end{align}
the 2-point function becomes
\begin{align}
\rho(x_1,x_2 | z ) = \dfrac{x_1 z}{(1 - x_1^2 z^2)(1 - x_1 x_2 z^2)}
\end{align}
the 3-point function becomes
\begin{align}
\rho(x_1,x_2,x_3 | z ) = \frac{zx_1 (1-qtx_1^4x_2^2z^6)\Big(1+z(x_1+x_2)(1+x_1x_2z^2) +
x_1^2x_2^2z^4\Big)}
{(1-x_1^3z^3)(1-qx_1^2x_2z^3)(1-x_1^3x_2^3z^6)(1-x_1x_2x_3z^3)} \end{align}

In general, $\rho$ is a polynomial of degree $(r-1)(r-2)/2$ in $t$ which provides a kind of
"separation of variables" $q$ and $t$. In particular, the 4-point function consists of
four different terms. The dependence on $z$ can be easily restored by dimensional argument,
hence, we omit it from the formulas below.

\be
\rho(x_1,x_2,x_3,x_4) \ = \frac{(x_1 + x_1 x_2 x_3) (N_0 + N_1 t + N_2 t^2 + N_3 t^3)}{D(x_1,x_2,x_3,x_4)}
\ee
where
\be
D(x_1,x_2,x_3,x_4) \ = (1 - x_{1}^2) (1 - q x_{1} x_{2}) (1 - x_{1} x_{2}) (1 - q^2 x_{1}^3 x_{2})
\times\nonumber\\\times  (1 - x_{1}^2 x_{2}^2 x_{3}^2)
(1 - q^2 x_{1}^2 x_{2} x_{3}) (1 - q x_{1}^2 x_{2} x_{3}) (1 - q^2 x_{1}^3 x_{2}^3 x_{3}^2)(1 - x_{1} x_{2} x_{3} x_{4})
\nonumber
\ee
\begin{align}
N_0 = (1-q^2 x_{1}^3 x_{2}^2 x_{3}) (1-q^2 x_{1}^5 x_{2}^3 x_{3}^2)\nonumber
\end{align}
\begin{align}
N_1 \ = \ & q x_{1} x_{2}-q x_{1}^3 x_{2}-q^2 x_{1}^3 x_{2}^2 x_{3}-q^3 x_{1}^4 x_{2}^2-q x_{1}^3 x_{2}^3 x_{3}^2
+q^2 x_{1}^4 x_{2}^3 x_{3}+q^3 x_{1}^5 x_{2}^3
-q^3 x_{1}^4 x_{2}^4 x_{3}^2+
q^4 x_{1}^5 x_{2}^4 x_{3}+
\nonumber \\ &
+q^3 x_{1}^5 x_{2}^5 x_{3}^2+q^3 x_{1}^6 x_{2}^3 x_{3}^3+q^3 x_{1}^6 x_{2}^4 x_{3}^2
-q^4 x_{1}^7 x_{2}^4 x_{3} + q^5 x_{1}^7 x_{2}^4 x_{3}^3 -q^4 x_{1}^8 x_{2}^4 x_{3}^2
-q^4 x_{1}^7 x_{2}^6 x_{3}^3
- q^3 x_{1}^8 x_{2}^5 x_{3}^3-
\nonumber \\ &
-q^5 x_{1}^8 x_{2}^6 x_{3}^2-q^5 x_{1}^9 x_{2}^4 x_{3}^3-q^4 x_{1}^8 x_{2}^6 x_{3}^4+q^5 x_{1}^9 x_{2}^5 x_{3}^4 -
q^5 x_{1}^9 x_{2}^6 x_{3}^5
+q^6 x_{1}^{11} x_{2}^7 x_{3}^4
+\nonumber \\ &
+[2]_qq^2x_1^4x_2^3x_3\Big(- q^{4} x_1^{5}x_2^{3}x_3^{2} ( -1+x_2x_1^{3}x_3^{2} ) +
 q^{2}x_1^{3}x_2 ( 1+x_2^{2}x_3^{2} )
 ( 1+x_1x_3 )+  qx_1^{3}x_2x_3^{2}-   ( 1+x_1x_3 ) \Big)-
\nonumber \\ &
-[3]_q qx_{1}^4 x_{2}^2 x_{3}^2
\nonumber
\end{align}
\begin{align}
N_2 \ = \ & q^2 x_{1}^3 x_{2}^2 x_{3}-q^3 x_{1}^5 x_{2}^3+q^3 x_{1}^5 x_{2}^4 x_{3} -
q^4 x_{1}^6 x_{2}^3 x_{3}-q^3 x_{1}^5 x_{2}^5 x_{3}^2-q^3 x_{1}^6 x_{2}^3 x_{3}^3
- q^5 x_{1}^6 x_{2}^4 x_{3}^2-q^4 x_{1}^7 x_{2}^3 x_{3}^2
- q^4 x_{1}^6 x_{2}^5 x_{3}^3+
\nonumber \\ &
+q^3 x_{1}^7 x_{2}^5 x_{3}^2
-q^4 x_{1}^7 x_{2}^5 x_{3}^4 +
q^5 x_{1}^8 x_{2}^5 x_{3}^3+q^5 x_{1}^8 x_{2}^6 x_{3}^2+q^5 x_{1}^9 x_{2}^4 x_{3}^3 +
q^4 x_{1}^9 x_{2}^5 x_{3}^4
-q^5 x_{1}^{10} x_{2}^5 x_{3}^3+q^5 x_{1}^9 x_{2}^6 x_{3}^5
+
\nonumber \\ &
+q^6 x_{1}^{10} x_{2}^6 x_{3}^4
-q^7 x_{1}^{11} x_{2}^6 x_{3}^3-q^5 x_{1}^{10} x_{2}^7 x_{3}^5
-q^6 x_{1}^{11} x_{2}^7 x_{3}^4 -
q^7 x_{1}^{11} x_{2}^8 x_{3}^5+q^7 x_{1}^{13} x_{2}^8 x_{3}^5+
\nonumber \\ &
+[2]_qqx_1^2x_2^2\Big((-1+x_2x_1^3x_3^2)+q^2x_1^4x_3x_2^2(1+x_2^2x_3^2)(1+x_1x_3)+
q^3x_2^2x_3^2x_1^5
-q^4x_1^7x_2^4x_3^3(1+x_1x_3)\Big)-
\nonumber \\ &
-[3]_q q^5x_{1}^{10} x_{2}^7 x_{3}^3
\nonumber
\end{align}
\begin{align}
N_3 = - q^4 x_{1}^6 x_{2}^4 x_{3}^2 (1-q^2 x_{1}^3 x_{2}^2 x_{3}) (1-q^2 x_{1}^5 x_{2}^3 x_{3}^2)
\nonumber
\end{align}

\section{Conclusion}

We described what we believe is an important parametrization of
the coefficients $c_Q^{(m,r)}$ in the MacDonald expansion (\ref{tspexpan})
of the extended superpolynomials of ref.\cite{DMMSS} for the torus knots.

The following table shows the structure of our calculation:
\be
P_{m,mk+r}(A) = \sum_{Q\vdash m} c_Q^{(m,r)}q^{-k\nu(Q^T)}t^{k\nu(Q)}M_Q^*(A)
\label{a1}
\ee
$$
\downarrow
$$
\be
{\cal P}_{m,mk+r}\{p\} = \sum_{Q\vdash m} c_Q^{(m,r)}q^{-k\nu(Q^T)}t^{k\nu(Q)}M_Q\{p\}
= e^{k\hat W} {\cal P}_{m,r}\{p\}
\label{a2}
\ee
$$
\downarrow
$$
\be
{\cal P}_{m,r}(p) = \sum_{Q\vdash m} c_Q^{(m,r)}M_Q\{p\} =
\label{a3}  \sum_{\stackrel{Q\vdash m}{l(Q)\leq r}} h_Q^{(m,r)}L_Q\{p\}
\ee
$$
\downarrow
$$
\be
P_{m,r}(A) = \sum_{\stackrel{Q\vdash m}{l(Q)\leq r}} h_Q^{(m,r)}L_Q^*(A)
= A^{r-m}P_{r,m}(A) =
 \label{a4}
\sum_{Y\vdash r} c_Y^{(r,r')}q^{-k'\nu(Y^T)}t^{k\nu(Y)} A^{r-m}M_Y^*(A)
\ee
The ordinary torus superpolynomial (\ref{a1}) is expanded,
as a function of $A$, into the MacDonald dimensions $M_Q^*(A)$.
For $m\geq 4$ this expansion becomes ambiguous
(the ambiguity is absent for the torus HOMFLY polynomials,
where only the hook diagrams $Q=[m-i,\underbrace{1,\ldots,1}_i]$
contribute).
The extended superpolynomial (\ref{a2}) is no longer ambiguous,
but instead it is not an invariant, it depends on the
choice of the braid representation.
Still, we believe that {\it it} is the central object to
study in the refined Chern-Simons theory \cite{rCS,ASh}.
For the torus knots, the extended superpolynomial (\ref{a2})
comes directly in the $W$-representation \cite{Wrep},
which explicitly describes "evolution" in the $k$-parameter,
so that the only unknown piece of the answer is
the "initial" condition at $k=0$, i.e.
the extended superpolynomial (\ref{a3}) of the knot
$T[m,r]$ with $r<m$.
If reduced back to the ordinary superpolynomials,
$P_{m,r}(A)$ is related by the obvious duality
$T[m,n]=T[n,m]$ to another superpolynomial, $P_{r,m}(A)$,
with less strands $r<m$ than the original one.
Accordingly, the sum over $Q$ for
${\cal P}_{m,r}\{p\}$ is actually "smaller" than for $k\neq 0$.
As an explicit manifestation of this fact, there is a special basis,
where the only contributing $Q$, while still having the size $m$,
have no more than $r$ rows, $l(Q)\leq r$.
It turns out that this special basis is formed not by
MacDonald, but by the simpler, Hall-Littlewood polynomials
$L_Q\{p\}$ (the same property persists in the basis of
"rescaled" Schur functions, but the expansion coefficients
are still more sophisticated in this case).

Note that we have discussed so far only the Hall-Littlewood decomposition of the
$(m,r)$ superpolynomials with $r < m, m\perp r$ and the "initial" condition given
at $k=0$. However, as we already mentioned there is a duality symmetry
$T[m,-n] \leftrightarrow T[m,n]$,
i.e. $T[m,r] \leftrightarrow T[m,m-r]$ and, simultaneously, $q\leftrightarrow
q^{-1}$, $t\leftrightarrow t^{-1}$. It allows one to describe also the case $k=-1$.
Namely, in addition to (\ref{dec}), one now has
\be
{\cal P}_{m,r-m}\{p\} = \sum_{Q\vdash m} c_Q^{(m,r)} q^{k\nu(Q^T)}t^{-k\nu(Q)}  M_Q\{p\} = \sum_{\stackrel{Q\vdash m}{l(Q)\leq r}} {\widetilde h}_Q^{(m,r)} {\widetilde L}_Q\{p\}
\ee
where ${\widetilde L}_Q$ and ${\widetilde h}_Q$ are the dual Hall-Littlewood
polynomials and coefficients, respectively:
\be\label{dual}
{\widetilde L}_Q\{p\} = L_Q\{p\} \Big|_{t=t^{-1} }\ ,
\ \ \ \ \ \ \ \ \ \ \ \ \ \ \ \ \ \ \ \  {\widetilde h}_Q^{(m,r)} = q^{m(m-1)/2} \cdot h_Q^{(m,m-r)}
\Big|_{{t=t^{-1}}\atop{q=q^{-1}}}
\ee

Further comments, explanations and examples (which substantially extend
the original list in \cite{DMMSS}) will be presented elsewhere.
We believe that appearance of two different expansion bases:
one for the $n$-evolution, another one for its "initial" condition
(i.e. for the $m$-evolution) is not an accident, and reflects some
additional duality structure.
Even more interesting is that this second evolution involves
only the Hall-Littlewood expansion, what implies a possible
existence of still another deformation of the entire construction,
perhaps, going as far as the Kerov-Askey-Wilson character expansion.
The situation here looks reminiscent of the one with
double elliptic deformations of integrable systems \cite{DELL}.

Extension of these results to other representations
(colored superpolynomials) looks straightforward
for the torus knots, but remains to be done.
Extension to similar series of {\it non}-torus knots
({\it a la} \cite{knMMM12}) is a far more interesting,
though a little less trivial exercise.
Last but not least, a hidden algebraic (integrable) structure
behind the extended superpolynomials (again, {\it al la} \cite{knMMM12})
is to be revealed.

\section*{Acknowledgements}

We are indebted to Ivan Cherednik, Petr Dunin-Barkovsky, Eugene Gorsky,
Sergei Gukov and Alexei Oblomkov for very useful and inspiring communications.
They also kindly compared some implications of our general formulas
with their own answers, obtained by different methods.
We understand that Alexey Oblomkov recently found an alternative effective
description of all the coefficients $c_Q^{(m,r)}$.

Our work is partly supported by Ministry of Education and Science of
the Russian Federation under contract 14.740.11.0608, by the Dynasty Foundation,
by RFBR grant 10-01-00536 and
by joint grants 11-02-90453-Ukr, 09-02-93105-CNRSL, 12-02-91000-ANF,
12-02-92108-Yaf-a, 11-01-92612-Royal Society.
The work of Sh.Shakirov is supported in part by Berkeley Center for Theoretical
Physics, by the National Science Foundation (award number 0855653), by the
Institute for the Physics and Mathematics of the Universe, by the US Department of
Energy under Contract DE-AC02-05CH11231.


\newpage

\begin{thebibliography}{12}

\bibitem{DMMSS} P.Dunin-Barkowski, A.Mironov, A.Morozov, A.Sleptsov and A.Smirnov,
arXiv:1106.4305
\bibitem{Kh} M.Khovanov and L.Rozhansky, Fund.
Math. {\bf 199} (2008) 1, math.QA/0401268; Geom. Topol. {\bf 12}
(2008) 1387, math.QA/0505056
\bibitem{superpols}
S.Gukov, A.Schwarz and C.Vafa, Lett.Math.Phys. {\bf 74} (2005) 53-74,
hep-th/0412243\\
N.M.Dunfield, S.Gukov and J.Rasmussen, Experimental Math. 15 (2006) 129-159,
math/0505662\\
S.Gukov and J.Walcher, hep-th/0512298\\
S.Gukov, A.Iqbal, C.Kozcaz and C.Vafa, arXiv:0705.1368\\
S.Gukov,     arXiv:0706.2369\\
N.Carqueville and D.Murfet, arXiv:1108.1081\\
S.Gukov and M.Stosic, arXiv:1112.0030
\bibitem{MMN} A.Mironov, A.Morozov and S.Natanzon, Theor.Math.Phys. 166 (2011) 1-22,
arXiv:0904.4227; Journal of Geometry and Physics, {\bf 62} (2012) 148-155
arXiv:1012.0433
\bibitem{Wrep} A.Morozov and Sh.Shakirov,
JHEP {\bf 0904} (2009) 064, arXiv:0902.2627; Mod.Phys.Lett. {\bf A24} (2009) 2659-2666,
arXiv:0906.2573;\\
G.Borot, B.Eynard, M.Mulase and B.Safnuk, arXiv:0906.1206\\
A.Alexandrov,
arXiv:1005.5715, arXiv:1009.4887
\bibitem{chi} M.Rosso and V.F.R.Jones, J. Knot Theory Ramifications, {\bf 2} (1993) 97-112\\
J.M.F.Labastida and M.Marino,
J.Knot Theory Ramifications, {\bf 11} (2002) 173\\
X.-S.Lin and H.Zheng, Trans. Amer. Math. Soc. 362 (2010) 1-18, math/0601267\\
A.Brini, B.Eynard and M.Mari\~no, arXiv:1105.2012
\bibitem{knMMM12}
A.Mironov, A.Morozov and And.Morozov, arXiv:1112.2654; arXiv:1112.5754
\bibitem{betadefo} A.Morozov, arXiv:1201.4595
\bibitem{EG}
E.Gorsky, arXiv:1003.0916
\bibitem{Obl}
A.Oblomkov, J.Rasmussen and V.Shende, arXiv:1201.2115 (with an Appendix
by Eugene Gorsky)
\bibitem{ASh} M.Aganagic and Sh.Shakirov, arXiv: 1105.5117
\bibitem{Che} I.Cherednik, arXiv:1111.6195
\bibitem{ShSh} Sh.Shakirov, arXiv:1111.7035\footnote{
A remarkably simple suggestion for the $\gamma$-factors in this paper
does not seem literally applicable to the torus knots,
it does not even reproduce the HOMFLY polynomials.
It is so attractive, however, and it is non-trivial that
such a simple ansatz provides polynomials with positive integer
coefficients. Therefore, there should be a prominent place for it
in the future theory of superpolynomials.
}
\bibitem{DELL}
H.Braden, A.Marshakov, A.Mironov and A.Morozov,
Nucl.Phys., {\bf B573} (2000) 553, hepth/9906240\\
see also a brief review in:
A.Mironov, Theor.Math.Phys., {\bf 129} (2001) 1581-1585,
hep-th/0104253\\
A.Mironov and A.Morozov, Phys.Lett.,
{\bf B475} (2000) 71-76, hepth/9912088; hepth/0001168\\
H.Braden, A.Gorsky, A.Odesskii and V.Rubtsov, hep-th/0111066
\bibitem{rCS} A.Iqbal, C.Kozcaz and C.Vafa, JHEP {\bf 0910} (2009) 069,
hep-th/0701156\\
H.Awata and H.Kanno, Int.J.Mod.Phys. {\bf A24} (2009) 2253-2306, arXiv:0805.0191


\end{thebibliography}
\end{document}